\documentclass{PoS}
\newcommand{\Z}{{\mathbb{Z}}}

\title{The decay of unstable strings in $SU(2)$ Yang-Mills theory}

\ShortTitle{The decay of unstable strings in $SU(2)$ Yang-Mills theory}

\author{\speaker{M. Pepe}\\
    INFN, Istituto Nazionale di Fisica Nucleare, Sezione di Milano-Bicocca \\
Edificio U2, Piazza della Scienza 3 - 20126 Milano, Italy.\\
        E-mail: \email{pepe@mib.infn.it}}

\author{U.-J.\ Wiese\\
        Albert Einstein Center for Fundamental Physics,
Institute for Theoretical Physics, Bern University,
Sidlerstr.\ 5, 3012 Bern, Switzerland.\\
        E-mail: \email{wiese@itp.unibe.ch}}

\abstract{We investigate the stability of strings connecting charges $Q$ in the
  representation $\{2Q+1\}$ of $SU(2)$ Yang-Mills theory in $(2+1)$ dimensions.  While the
  fundamental $\{2\}$-string between two charges $Q = \frac{1}{2}$ is unbreakable and
  stable, the string connecting static charges transforming under any other representation
  $Q \geq 1$ is unstable and decays. A charge $Q = 1$ can be completely screened by gluons
  and so the adjoint $\{3\}$-string ultimately breaks. A charge $Q=\frac{3}{2}$ can be
  only partially screened to a fundamental charge $Q=\frac{1}{2}$. Thus, stretching a
  $\{4\}$-string beyond a critical length, it decays into the stable $\{2\}$-string by
  gluon pair creation. The complete breaking of a $\{5\}$-string happens in two steps, it
  first decays into a $\{3\}$-string and then breaks completely. A phenomenological
  constituent gluon model provides a good quantitative description of the energy of the
  screened charges at the ends of an unstable string.}

\FullConference{The XXVII International Symposium on Lattice Field Theory\\
                 July 26-31, 2009\\
                 Peking University, Beijing, China}

\begin{document}

\section{Introduction}
At low temperatures quarks are confined inside hadrons. As one pulls apart a
quark-anti-quark pair, its energy increases linearly with the distance. At some point the
energy stored in the flux tube is sufficient to pop a quark-anti-quark pair out of the
vacuum. In this way the string breaks or, alternatively, the string state decays into a
meson pair \cite{Boc90,Kne98,Phi99,Gli05,Bal05}. As one increases the quark mass, the
energy that needs to be stored in the flux tube in order to create a quark-anti-quark pair
from the vacuum will be larger. This means that the sources have to be pulled apart to
larger distances. As one sends the quark mass to infinity, the string decay scale goes to
infinity as well, and one recovers the Yang-Mills theory where quark
degrees of freedom have been removed from the dynamics. On the other hand, the string
connecting two adjoint sources is still unstable due to pair-creation of dynamical
gluons. This effect has been investigated in
\cite{Pou97,Ste99,Phi99a,Kra03,For00,Kal02,Gre07}.

Observing the decay of unstable strings and investigating the characteristics of this
process provide valuable insights into the physics of confinement. The property of
$N$-ality is important to understand the decay of unstable strings in $SU(N)$ Yang-Mills
theory. In this theory, the only dynamical degrees of freedom are gluons that transform
under the adjoint representation of the gauge group. Since the center subgroup of $SU(N)$
is $\Z(N)$, the representations split into $N$ different $N$-ality sectors. Starting from
a given representation one can reach all other representations in the same $N$-ality
sector by coupling the initial representation with an arbitrary number of adjoint
representations. On the other hand, by coupling a given representation to an arbitrary
number of adjoint representations, one can never reach another $N$-ality sector.
As a physical consequence, $N$-ality implies that, by gluon emission, a given source
representation can only be screened to representations in the same $N$-ality sector. If
two representations belong to different sectors, this cannot happen. In every $N$-ality
sector there is one stable string, i.e. the one with the minimal string tension. All
other strings in that sector decay into the stable one for a sufficiently
large distance between the sources.

For simplicity, we study the dynamics of strings in $(2+1)$-d $SU(2)$ Yang-Mills theory
which has the center $\Z(2)$. We expect that other theories in $(3+1)$ dimensions or with
other gauge groups show similar behavior. We consider strings connecting two static
charges $Q$ in the representation $\{2Q+1\}$ of $SU(2)$. We denote them as
$\{2Q+1\}$-strings which should not be confused with $k$-strings. When $Q$ is an integer,
the $\{2Q+1\}$-strings are unstable and they eventually break at large distances. On the
other hand, when $Q$ is a half-integer, the $\{2Q+1\}$-strings are still unstable but
ultimately unbreakable. At sufficiently large distances, all these strings have the same
tension $\sigma_{1/2}$ given by the fundamental $\{2\}$-string. Dynamical gluons screen
the static charges $Q$ at the two endpoints of the $\{2Q+1\}$-string: when a gluon pair
pops out of the vacuum, the external sources $Q$ are screened and they behave as sources
$Q-1$. Hence, the $\{2Q+1\}$-string decays to a $\{2Q-1\}$-string and its tension is
abruptly reduced \cite{Arm03,Gli05a}. Here, using the multi-level simulation technique of
\cite{Lue04}, we present results of a detailed study of string decay \cite{Pep09}. Some
indications for string decay were presented in \cite{Del03}.

\section{The numerical study}
We perform numerical simulations of $SU(2)$ Yang-Mills theory on a cubic lattice in
$(2+1)$ dimensions. We consider the standard Wilson action given by the path-ordered product
of the link variables in the fundamental representation along an elementary plaquette. The
observable we measure is the two-point function of Polyakov loops $\Phi_Q(x)$ in the
$\{2Q+1\}$ representation. In this way, we insert external color charges $Q$ into the
system and the corresponding potential $V_Q(r)$ is extracted from
\begin{equation}
\langle \Phi_Q(0) \Phi_Q(r) \rangle \sim \exp(- \beta V_Q(r)).
\end{equation}
The numerical simulations have been performed at an inverse temperature as large as $\beta =
64$ in lattice units in order to enhance the projection on the ground state of the string. The
lattice size in the spatial direction was $L = 32$. We run at bare gauge coupling $4/g^2 =
6.0$: this puts the deconfinement phase transition at $\beta_c \approx 4$.  Although we
consider a moderate coupling, we expect that discretization effects are marginal and that our
results stay unchanged, at least qualitatively, in the continuum limit.  The measurements
of the Polyakov loop correlators span a wide range of values, from $10^{-8}$ to
$10^{-135}$: this was possible thanks to the very powerful multi-level simulation
technique developed by L\"uscher and Weisz \cite{Lue04}. We have slightly improved this
method by slicing the lattice not only into slabs in time, but also into blocks in space.
After an elaborate tuning of the parameters of the multi-level algorithm, we have measured
the potentials $V_Q(r)$ for the $\{2\}$-, $\{3\}$-, $\{4\}$-, and $\{5\}$-strings. In
figure 1-left, we observe the decay of the $\{4\}$-string to the $\{2\}$-string at $r
\approx 8$ with a sudden reduction of the tension down to the value of the fundamental
string. In figure 1-right, the $\{5\}$-string has a first decay at distance $r
\approx 6$, reducing its tension to the one of the adjoint $\{3\}$-string. Then, at $r
\approx 10$, the string breaks completely, at about the same distance as the adjoint
$\{3\}$-string. Consistent with expectations, the tension of a string connecting two
charges $Q$ is the same, no matter whether those charges are screened or not.

\begin{figure}[h]
\hskip-.5cm
\begin{minipage}[t]{7.5cm}
\begin{center}
\includegraphics[width=1.04\textwidth]{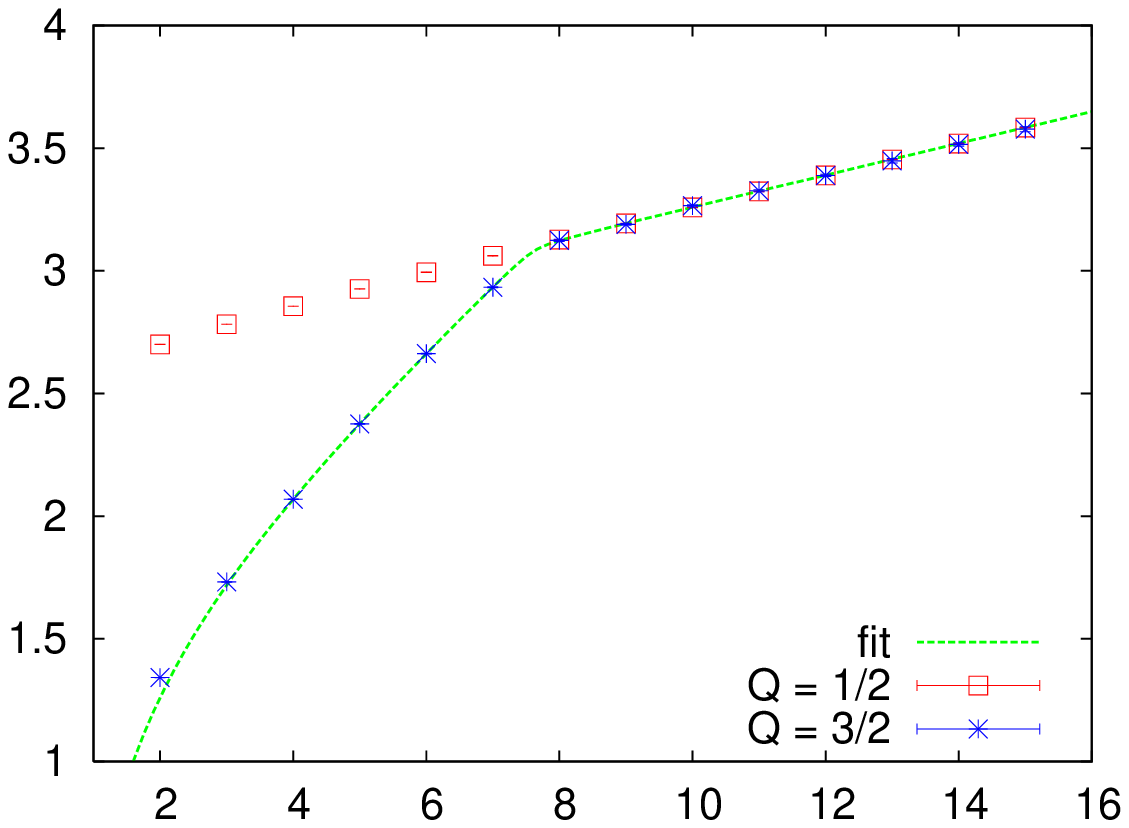}
\end{center}
\end{minipage}
\hfill
\begin{minipage}[t]{7.5cm}
\begin{center}
\includegraphics[width=1.04\textwidth]{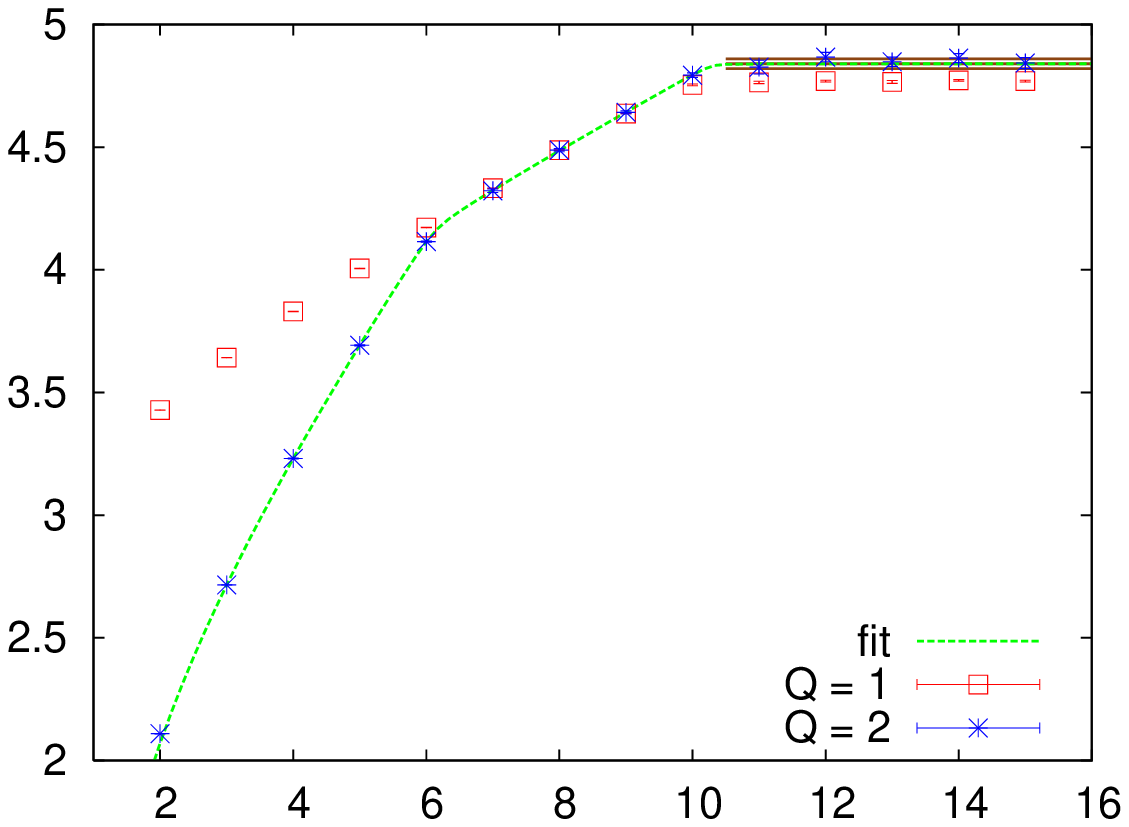}
\end{center}
\end{minipage}
\caption{\it Left: Potential $V_Q(r)$ of two static color charges with half-integer
  charges $Q = \frac{1}{2}$ and $Q = \frac{3}{2}$. For a more convenient comparison of the
  slopes, the $Q = \frac{1}{2}$ data have been shifted by a constant. Right: The same for
  $Q = 1$ and $Q = 2$. The lines are the fits of the Monte Carlo data obtained using the
  multi-channel model. The horizontal band at $2 M_{0,2}=4.84(2)$ corresponds to twice the
  mass of a source of charge $Q=2$. This value has been obtained from the measurement of a
  single Polyakov loop.}
\end{figure} 

In order to fit the static potential connecting two fundamental charges, we use
\cite{Lue80,Lue81,Lue04a}
\begin{equation}
V_{1/2}(r) = \sigma_{1/2} r - \frac{\pi}{24 r} + 2 M + {\cal O}(1/r^3),
\end{equation}
The quality of the fit is very good and we measure the string tension $\sigma_{1/2} =
0.06397(3)$.  Furthermore, the Monte Carlo data show an excellent agreement with the
coefficient $- \frac{\pi}{24}$ of the L\"uscher term. The constant term, describing the
``mass'' contribution of an external charge $Q = \frac{1}{2}$ to the total energy, is
given by $M = 0.109(1)$. However, due to ultra-violet divergences, this ``mass'' itself is not
physical. 

The energy scale, $\Lambda_{\mbox{QCD}}$, of the Yang-Mills theory is not well separated
from the typical distances of the string decays. Thus, unlike the string behavior at
asymptotic distances, one cannot describe the string decay in a fully systematic
low-energy effective string theory. Moreover, unlike the tension $\sigma_{1/2}$ of the
stable fundamental string, the tension $\sigma_Q$ of an unstable $\{2Q+1\}$-string (with
$Q \geq 1$) cannot be defined unambiguously. We define $\sigma_Q$ as a
fit parameter of the Monte Carlo data to a simple phenomenological model. In this
model, the $\{2Q+1\}$-string is described as a multi-channel system.

In the phenomenological multi-channel model, the energy of a $\{2Q+1\}$-string connecting
two charges $Q$, which results from the screening of a larger charge $Q+n$ by $n$ gluons,
is given by
\begin{equation}
E_{Q,n}(r) = \sigma_Q r - \frac{c_Q}{r} + 2 M_{Q,n}.
\end{equation}
In general, $c_Q$ is the coefficient of a sub-leading $1/r$ correction: this term does not
necessarily assume the asymptotic L\"uscher value $- \frac{\pi}{24}$. We denote the
``mass'' that describes the contribution of an original charge $Q+n$ that $n$ gluons have
screened down to the value $Q$ by $M_{Q,n}$. Like the ``mass'' $M = M_{1/2,0}$,
the ``masses'' $M_{Q,n}$ themselves are not physical due to ultra-violet divergent
contributions. On the other hand, the mass differences $\Delta_{Q,n} = M_{Q-1,n+1} -
M_{Q,n}$ have a physical meaning since the divergent pieces cancel.  The two-channel
Hamiltonians, $H_1$ and $H_{3/2}$, describe the $\{3\}$- and $\{4\}$-strings; the
$\{5\}$-string is described by the three-channel Hamiltonian $H_2$
\begin{eqnarray}
&&H_1(r) = \left(\begin{array}{cc} E_{1,0}(r) & A \\ A & E_{0,1}(r)
\end{array}\right), \nonumber \\
&&H_{3/2}(r) = \left(\begin{array}{cc} E_{3/2,0}(r) & B \\ B & E_{1/2,1}(r)
\end{array}\right), \nonumber \\
&&H_2(r) = \left(\begin{array}{ccc} E_{2,0}(r) & C & 0 \\ C & E_{1,1}(r) & A \\
0 & A & E_{0,2}(r) \end{array}\right).
\end{eqnarray}
The parameters $A$, $B$, and $C$ are decay amplitudes --- which we assume to be
$r$-independent --- and the potential $V_Q(r)$ is the energy of the ground state of
$H_Q$. Using the multi-channel model, in Figure 2 we compare the forces $F(r) = -
dV(r)/dr$ for the $\{2\}$-, $\{3\}$-, $\{4\}$-, and $\{5\}$-string cases with the results
of the numerical simulations. 

In table 1-left, we list the tensions $\sigma_Q$ determined by a fit of the Monte Carlo
data: the simple multi-channel model works rather well. Interestingly, the ratios
$\sigma_Q/\sigma_{1/2}$ do not obey the conjectured Casimir scaling \cite{Ole81,Amb84},
i.e.\ they are not equal to $4 Q(Q+1)/3$. In table 1-right we list the ``masses''
$M_{Q,n}$. It is important to note that, within the error bars, the mass differences
$\Delta_{Q,0} = M_{Q-1,1} - M_{Q,0}$ all take the same value $M_G = 0.65(5)$, independent
of $Q$. According to this result, $M_G$ can be interpreted as a constituent gluon mass: in
units of the fundamental string tension, it takes the value $M_G/\sqrt{\sigma_{1/2}} =
2.6(2)$. In contrast to the string tension, $M_G$ is not unambiguously defined. Again, it is
obtained from the fit parameters using the phenomenological multi-channel model. The mass
difference $\Delta_{1,1} = M_{0,2} - M_{1,1} = 0.71(3)$ shows that the dynamical
creation of a second constituent gluon has an energy cost slightly larger than
$M_G$. Finally, it is interesting to note that the mass of two constituent gluons $2 M_G =
1.3(1)$ is compatible with the $0^+$ glueball mass $M_{0^+} = 1.198(25)$ measured at the
same value of the bare coupling \cite{Mey03}.  The constituent gluon mass $M_G$ is also
related to the distance scale for string decay and string breaking. In fact, the distance
at which the $\{4\}$-string decays into the $\{2\}$-string is $r \approx 2
M_G/(\sigma_{3/2} - \sigma_{1/2}) = 7.3(6)$. Similarly, the string breaking distance of
the $\{3\}$- and of the $\{5\}$-string is given by $r \approx 2 M_G/\sigma_1 = 9.0(7)$.

\begin{figure}[h]
\hskip-.5cm
\begin{minipage}[t]{7.5cm}
\begin{center}
\includegraphics[width=1.04\textwidth]{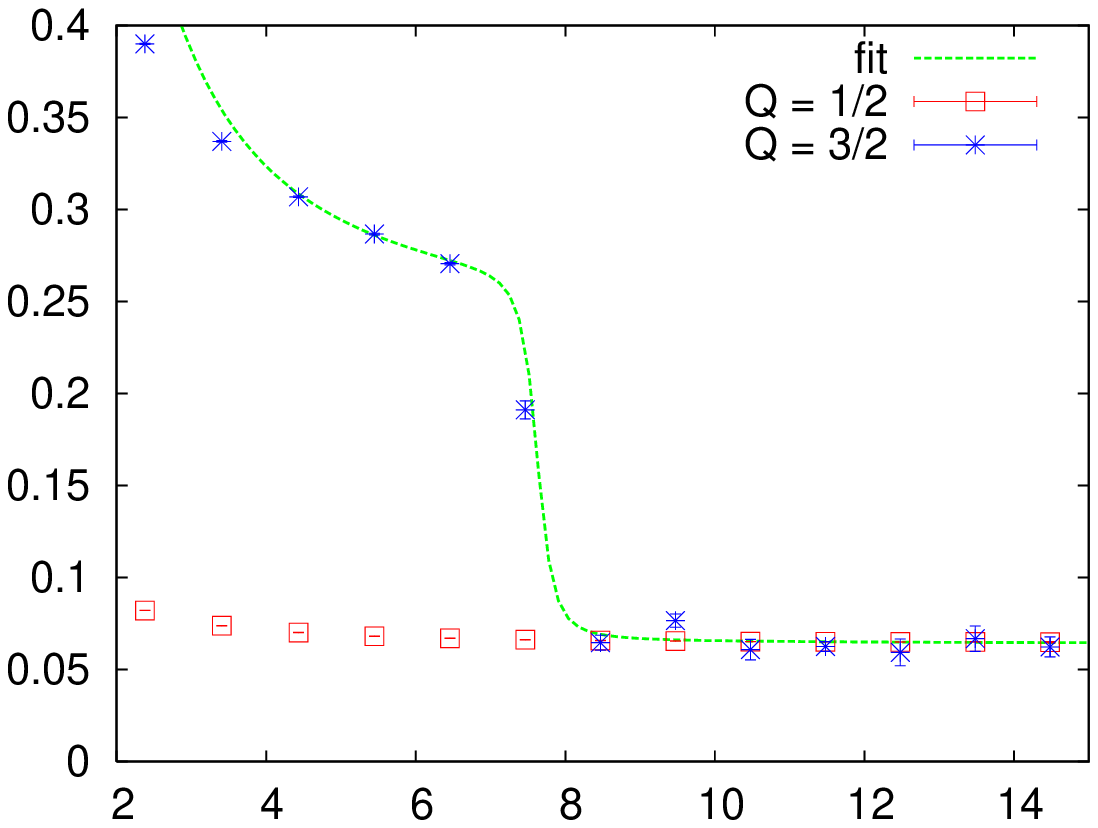}
\end{center}
\end{minipage}
\hfill
\begin{minipage}[t]{7.5cm}
\begin{center}
\includegraphics[width=1.04\textwidth]{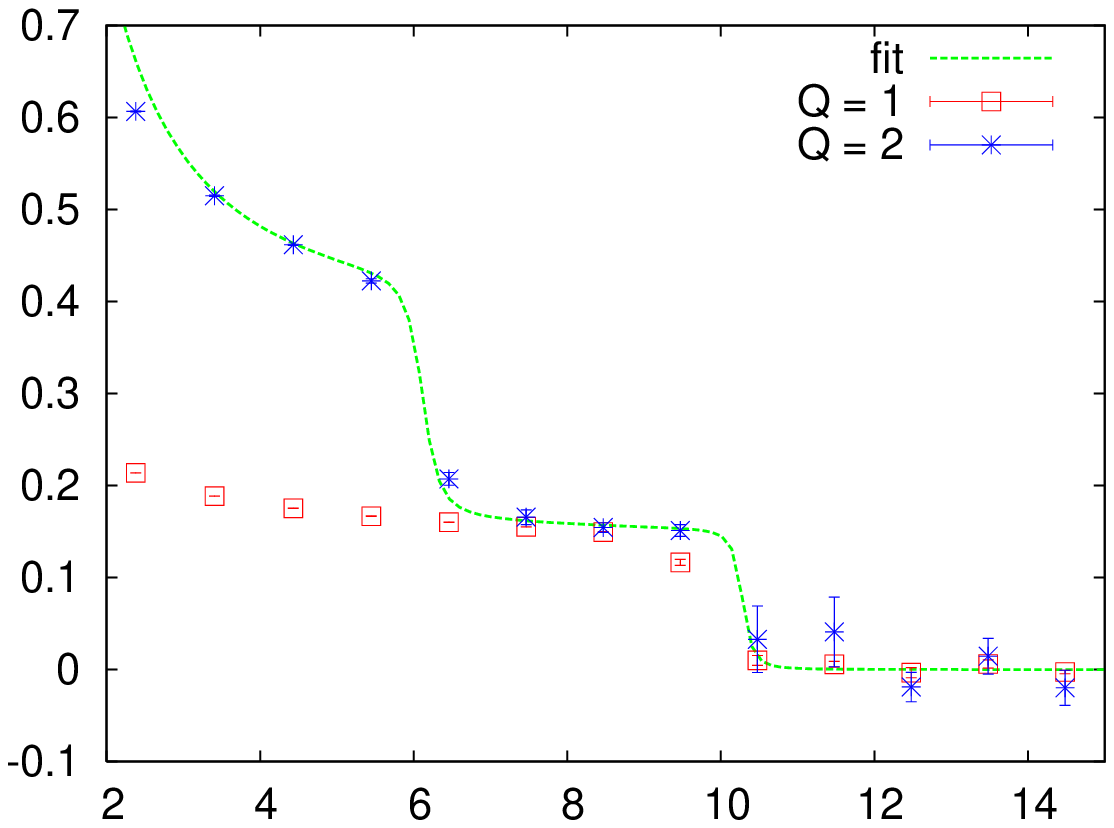}
\end{center}
\end{minipage}
\caption{\it Left: Forces $F(r)$ between the external charges $Q = \frac{1}{2}$ and $Q =
  \frac{3}{2}$. Right: The same for external charges $Q = 1$ and $Q = 2$. The lines are
  the fits of the Monte Carlo data using the multi-channel model.}
\end{figure}

\begin{table}[h]
{\small
\begin{minipage}[h]{6.5cm}
\begin{center}
\begin{tabular}{|c|c|c|c|}
\hline
$Q$ & $\sigma_Q$ & $\sigma_Q/\sigma_{1/2}$ & $4 Q(Q+1)/3$ \\
\hline
\hline
1/2 & 0.06397(3) & 1       & 1   \\ 
\hline
1   & 0.144(1)   & 2.25(2) & 8/3 \\ 
\hline
3/2 & 0.241(5)   & 3.77(8) & 5   \\ 
\hline
2   & 0.385(5)   & 6.02(8) & 8   \\ 
\hline
\end{tabular}
\end{center}
\end{minipage}
\hskip.1cm
\begin{minipage}[h]{6.5cm}
\begin{center}
\begin{tabular}{|c|c|c|c|c|c|}
\hline
$Q$ & $M_{Q,0}$ & $M_{Q-1,1}$ & $M_{Q-2,2}$ & $\Delta_{Q,0}$ & $\Delta_{Q-1,1}$ \\
\hline
\hline
1/2 & 0.109(1) & ---      & ---     & ---     & ---     \\ 
\hline
1   & 0.37(3)  & 1.038(1) & ---     & 0.67(3) & ---     \\ 
\hline
3/2 & 0.72(5)  & 1.32(5)  & ---     & 0.60(5) & ---     \\ 
\hline
2   & 1.04(3)  & 1.71(3)  & 2.42(1) & 0.67(3) & 0.71(3) \\ 
\hline
\end{tabular}
\end{center}
\end{minipage}
}
\caption{\it Left: Values of the string tensions $\sigma_Q$ obtained using the
  multi-channel model. The expected values of the ratio $\sigma_Q/\sigma_{1/2}$ assuming
  Casimir scaling ($4 Q(Q+1)/3$) is compared with the measurements obtained from numerical
  simulations. Right: Values of the ``mass'' $M_{Q,n}$ of an external charge $Q+n$
  screened by $n$ gluons to the value $Q$. The differences $\Delta_{Q,n} = M_{Q-1,n+1} -
  M_{Q,n}$ are also shown in the last two columns.}
\end{table}

\section{Discussion and outlook}
The process of strand rupture in a cable consisting of a bundle of strands is a classical
analog of the quantum string decay. Suppose that one stretches a cable further and
further: at some point individual strands eventually rupture, thereby abruptly reducing
the tension of the cable. However it is not clear whether the strand picture provides only
an intuitive analog or describes the actual anatomy of decaying $\{2Q+1\}$-strings. This
is a an interesting question that requires a detailed investigation of the internal
structure of $\{2Q+1\}$-strings.

In this paper, we have discussed the results of numerical simulations in $SU(2)$
Yang-Mills theory. It would be interesting to extend this study to other $SU(N)$ gauge
theories as well. For instance, $SU(4)$ Yang-Mills theory has two distinct unbreakable
strings due to its $\Z(4)$ center symmetry: the first stable $k$-string connects external
charges in the $\{4\}$ and $\{\overline 4\}$ representations, and the second stable
$k$-string connects two sources in the $\{6\}$-representation. For sources transforming
under larger representations with non-trivial $N$-ality, one then expects cascades of
string decays down to the $\{4\}$-string or down to the $\{6\}$-string. All other
representations belong to the trivial $N$-ality sector and strings in that sector
ultimately break completely.

The investigation of gauge groups other than $SU(N)$ is also interesting in many
respects. In particular, one expects new effects that are not present for $SU(N)$. For
instance, the groups $Sp(N)$ are simply connected and all have the same center
$\Z(2)$. The first group of this sequence is $Sp(1) = SU(2) = Spin(3)$; the next one is
$Sp(2) = Spin(5)$, which is also the universal covering group of $SO(5)$. In $Sp(2)$
Yang-Mills theory the only stable string is the one connecting two fundamental sources in
the $\{4\}$ representation. The adjoint representation is $\{10\}$ and its string breaks
by pair creation of gluons. However, in contrast to $SU(N)$ groups, $Sp(2)$ has a
center-neutral representation $\{5\}$ with a smaller size than the adjoint
representation. Since in $Sp(2)$
\begin{equation}
\{5\} \otimes \{10\} = \{5\} \oplus \{10\} \oplus \{35\},
\end{equation}
a single gluon can screen a charge in the representation $\{5\}$ only to a $\{10\}$ or a
$\{35\}$ representation, but not to a singlet. It is natural to expect that the unstable
$\{5\}$-string has a smaller tension than the adjoint string. Thus, although a charge in
the representation $\{5\}$ needs two adjoint charges to be completely screened, the string
should break in a single step by the dynamical creation of four gluons, without any
intermediate string decay.

The relevance of the center of the gauge group in the process of string decay is also an
important issue to be addressed. The simplest Lie group with a trivial center is the
exceptional group $G(2)$. Despite the triviality of the center, which implies that there
are no stable strings, $G(2)$ Yang-Mills theory confines color \cite{Hol03} and it has a
first order deconfinement phase transition \cite{Pep05,Pep07}. In fact, the order of the
deconfinement phase transition is not controlled by the center symmetry but by the size of
the gauge group \cite{Hol04}. In $G(2)$ Yang-Mills theory, the string connecting charges
in the fundamental $\{7\}$ representation is unstable. In fact, it ultimately breaks since
the $\{7\}$ representation can be completely screened by gluons in the adjoint $\{14\}$
representation. More precisely, since in $G(2)$
\begin{equation}
\{7\} \otimes \{14\} = \{7\} \oplus \{27\} \oplus \{64\},
\end{equation}
a single gluon can eventually screen a charge $\{7\}$ only to a $\{27\}$ or a
$\{64\}$. Casimir scaling has been observed for $G(2)$ Yang-Mills theory in
(3+1) dimensions, including the $\{27\}$- and the $\{64\}$-string \cite{Lip08}. Hence, the
string tensions of the $\{27\}$- and of the $\{64\}$-string are larger than the tension of
the $\{7\}$-string. This makes the $\{7\}$-string stable against decay due to the creation
of a single pair of gluons. A similar argument makes the $\{7\}$-string stable also
against decay due to the creation of two pairs of gluons. Since a $\{7\}$ charge can be
completely screened by three adjoint gluons, we expect that the fundamental $\{7\}$-string
has no intermediate decay and breaks in a single step by the simultaneous creation of
six gluons.

\acknowledgments 
M.P. acknowledges useful discussions with F.~Gliozzi and J.~Greensite.  This work is
supported in part by funds provided by the Schweizerischer Nationalfonds (SNF). The
``Albert Einstein Center for Fundamental Physics'' at Bern University is supported by the
``Innovations- und Kooperationsprojekt C-13'' of the Schweizerische
Uni\-ver\-si\-t\"ats\-kon\-fe\-renz (SUK/CRUS).

\end{document}